\documentclass[preprint,superscriptaddress,showpacs,floatfix,amsmath,amssymb]{revtex4}
\usepackage{color}
\usepackage{graphicx}
\begin{document}

%\preprint{APS/123-QED}

\title {Superconductivity in an extended Hubbard model with attractive interaction}
\author{E. J. Calegari}%
\affiliation{Laborat\'orio de Teoria da Mat\'eria Condensada,
Departamento de F\'{\i}sica - UFSM, 97105-900\\
 Santa Maria, RS, Brazil}%
\author{S. G. Magalh\~aes}
\affiliation{Instituto de F\'isica, Universidade Federal Fluminense\\  Av. Litor\^anea s/n, 24210,
346, Niter\'oi, Rio de Janeiro, Brazil}
\author{C. M. Chaves}%
\email{cmch@cbpf.br}
\author{A. Troper}%
\affiliation{Centro Brasileiro de Pesquisas F\'{\i}sicas, Rua Xavier Sigaud 150, 22290-180,
 Rio de Janeiro, RJ, Brazil}
%\%
\date{\today}

\begin{abstract}
In this work, a two-dimensional one-band Hubbard model is investigated within a two-pole approximation. The model presents a non-local attractive potential $U (U<0)$ that allows the study of
d-wave superconductivity and also includes hopping up to second-nearest-neighbors. 
The two-pole scheme has been proposed to improve the Hubbard-I approximation. The analytical results show a more complex form for the gap $\Delta(T)$, when compared to the one
obtained in the latter approximation. Indeed, new anomalous correlation functions associated with the superconductivity are involved in the calculation of $\Delta(T)$. Numerical results in a range of temperatures are presented. Moreover, the structure of the quasiparticle bands and the topology of the Fermi surface are studied in detail in the normal state. Connections with some experimental results are also included.

\end{abstract}


\maketitle

\section{Introduction}
Superconductivity in strongly correlated systems is a field with  plenty of challenging problems. Several 
non-usual properties of high temperature superconductors (HTSC) \cite{cuprates} still needed to be properly clarified. In particular, some experimental systems exhibit important deviations from the standard BCS theory. 
For instance, the superconductor gap behaviour as a function of temperature in some borocarbides display a 
non-monotonic feature at lower temperatures. In fact, such gap decreasing 
in 
$RNi_{2}B_{2}C (R=Dy,Ho,Er,Tm)$ and $ErNi_{2}B_{2}C$ \cite{borocarbides} 
when the temperature decreases towards to 
%zero
%as compared with the gap at temperature 
$T=0$. 
%clearly appears. 
This effect ascribed due to a competition between the superconductivity and  antiferromagnetic correlations, which are absent in a weak coupling BCS-like superconductivity, appear naturally from the formulation which is shown in 
this work. Another quite interesting feature of HTSC 
%a strongly %coupling superconductor 
systems 
is the behaviour found in some cuprates \cite{cuprates1}
in the low doping region, the so called pseudo-gap region, 
%is the appearence ¨%of 
which gives rise to an anomalous Fermi surface leading to a pseudo-gap. 
%In spite, 

In this very complex problem, a number of theories has been proposed in order to explain the presence of a pseudo-gap region \cite{cuprates}. 
%Within 
In the present formulation,  we claim that the appearance of a pseudo-gap 
%is due to 
can be ascribed 
%due a 
to a more detailed many-body treatment in which superconductor and AF correlations compete. 

Although BCS-like approach \cite{BCS}
has been widely used to describe 
these physical systems, it is well recognized that superconductivity is a two-dimensional problem in which strong correlations play a fundamental role \cite{correlations}. Thereby, we apply a two-pole approximation \cite{Roth69,Edwards95,Eleonir2005} to deal with the strong interaction coupling. Here we consider a d-wave symmetry gap and therefore, a non-local attractive interaction is used \cite{bib1,prbcaixeiro}.
The net attractive interaction $(U<0)$ may result, for example, from the elimination of the electron-phonon like coupling through a canonical transformation \cite{kittel} or, alternatively, from an electronic mechanism proposed by Hirsch \cite{Hirsch2} which
may produce, for a certain range of parameters, an effective attractive interaction.

In this work, we focused on the many-body renormalized normal state of these systems. Our obtained Fermi surface  is  consistent 
with recent claims in the literature \cite{cuprates1,singleton} about the presence of hole-pockets due to antiferromagnetic correlations. Moreover, we discuss  
some thermodynamical properties of the superconducting regime, 
namely the critical temperature $T_{c}$, the zero temperature superconducting gap $\Delta_{0}$ 
and temperature dependence of the gap $\Delta(T)$ for various 
dopings $\delta\equiv 1- n_{T}$ (with $n_T=n_{\sigma}+n_{-\sigma}$) and interaction $U$. The $n_{\sigma}$ represents the average occupation per site of electrons with spin $\sigma=\uparrow,\downarrow$.

This paper is organized as follows. In the section \ref{sec:II},  we present a general formulation describing the model as well as the ingredients 
of the normal state, e.g. the quasi-particle and the special characteristics of the Fermi surface (FS). In section \ref{sec:III}, we present the superconducting state which appears from the application of the two-pole approximation. In section \ref{sec:IV}, we exhibit self-consistent numerical results and conclusions for both the normal and the superconducting states . The Appendix \ref{app:A} briefly describes  the main points involved in the two-pole approach whereas in the Appendix \ref{app:B} the correlation functions involved in the Green's functions governing the superconducting and the normal states are displayed.

\section{General Formulation}
\label{sec:II}

The Hamiltonian studied here is, in a standard notation 
\begin{equation}
H=\sum_{\langle \langle ij \rangle\rangle \sigma} t_{ij}d_{i\sigma}^{\dag}d_{j\sigma}+
U\sum_{\langle ij \rangle \sigma} n_{i,\sigma}^{d} n_{j,-\sigma}^{d}-\mu\sum_{i\sigma}d_{i\sigma}^{\dag}d_{i\sigma}
\label{eqH}
\end{equation}
where $\langle \langle ...\rangle\rangle$ indicates the sum over the first and the second-nearest-neighbors of $i$ and $\mu$ is the chemical potential.
The two-dimensional dispersion relation is given by:
\begin{equation}
\varepsilon_{\vec{k}} = 2t(\cos(k_xa)+\cos(k_ya))+4t_2\cos(k_xa)\cos(k_ya).
\label{ed}
\end{equation}

In the present work, we adopted the two-pole approximation \cite{Roth69,Edwards95} which consists in choosing a 
set of operators describing the most important excitations of the system. The details of the method are given in Appendix \ref{app:A}.  In the present case, the set of operators considered is $\left\{d_{i,\sigma },n_{i,-\sigma }^{d}d_{i,\sigma },d_{i,-\sigma }^{\dagger},n_{i,\sigma }^{d}d_{i-\sigma }^{\dagger}\right\}$. The first two are associated with the normal state whereas the last two are associated with the superconductivity \cite{Edwards95,Eleonir2005}. 
Following the method exhibited in the Appendix \ref{app:A}, 
the one-particle Green's function for the normal state is:
\begin{equation}
G_{N\sigma}^{dd}(\vec{k},\omega)=\frac{Z_{1\vec{k}\sigma}}{\omega -\omega_{1\vec{k}\sigma}}+\frac{Z_{2\vec{k}\sigma}}{\omega -\omega_{2\vec{k}\sigma}}
\label{eqG11N}
\end{equation} 
with
\begin{equation}
Z_{1\vec{k}\sigma}=\frac{1}{2}+\frac{\overline{U}-2U_1-\varepsilon_{\vec{k}}+W_{\vec{k}\sigma}}{2X_{\vec{k}\sigma}},
\label{Z1}
\end{equation} 
\begin{equation}
Z_{2\vec{k}\sigma}=1-Z_{1\vec{k}\sigma}
\label{Z2}
\end{equation} 
and
\begin{equation}
\overline{U}=\frac{U_2+n_{-\sigma}(U_1-2U_2)}{n_{-\sigma}(1-n_{-\sigma})}.
\label{Ub}
\end{equation} 

The quasiparticle bands are:
\begin{equation}
\omega_{1\vec{k}\sigma}=\frac{\overline{U}+\varepsilon_{\vec{k}}-2\mu+W_{\vec{k}\sigma}}{2}
-\frac{X_{\vec{k}\sigma}}{2},
\label{w1}
\end{equation} 
%
%and
%
\begin{equation}
\omega_{2\vec{k}\sigma}=\omega_{1\vec{k}\sigma}+X_{\vec{k}\sigma}
\label{w2}
\end{equation} 
where
\begin{equation}
X_{\vec{k}\sigma}=\sqrt{(\overline{U}-\varepsilon_{\vec{k}}+W_{\vec{k}\sigma})^2+4U_1(\varepsilon_{\vec{k}}-W_{\vec{k}\sigma})+\widetilde{U}}
\label{Xk}
\end{equation} 
and
\begin{equation}
\widetilde{U}=\frac{4U_2(U_2-U_1)}{n_{-\sigma}(1-n_{-\sigma})}.
\label{Util}
\end{equation} 
The effective interactions $U_{1}$, $U_{2}$ and the band shift $W_{\vec{k}\sigma}$ 
are defined in the Appendix \ref{app:A} (see Eqs. (\ref{U_1})-(\ref{Apx32})).
Here, as we are assuming a paramagnetic state of a translationally invariant system,  
$\langle n_{i,\sigma}\rangle=\langle n_{i,-\sigma}\rangle=\langle n_{-\sigma}\rangle$. 
It should be noticed that, due to many body effects, 
%it appears 
in the pole structure of the Green's functions, in the normal paramagnetic phase,
%, in particular, 
there is a spin-spin correlation function which exhibit only antiferromagnetic (AF)  short range correlations. That is not in contradiction in our previous paramagnetic assumption.
Moreover, in order to simplify the notation, we write $\langle n_{-\sigma}\rangle=n_{-\sigma}$.

From the Green's function in equation (\ref{eqG11N}), we find the spectral function
\begin{equation}
A_{\sigma}(\vec{k},\omega)=-\frac{1}{\pi}\mbox{Im}[{G_{N\sigma }^{dd}(\vec{k},\omega)}].
\label{Awks}
\end{equation}

The Fermi surface is obtained from $A_{\sigma}(\vec{k},\omega=0)$.

\section{The superconducting state}
\label{sec:III}

In the superconducting state, the Green's function $G^{dd}$ is written as:

\begin{equation}
 G_{S\sigma}^{dd}(\vec{k},\omega)=\frac{A'(\omega)-(\omega+E_{11})(1+n_{-\sigma}^2\frac{U}{\theta})^2\Delta_{\vec{k}}^2}{P(\omega)}
\label{eqGS11}
\end{equation} 
where
\begin{equation}
A'(\omega)=\alpha_0+ \alpha_1\omega+ \alpha_2\omega^2+ \alpha_3\omega^3
\label{eqA}
\end{equation} 
with
\begin{eqnarray}
%\begin{align}
 \alpha_0=&(E_{12}^2-E_{11}E_{22})
%\nonumber\\ 
%& \times
(E_{22}-2n_{-\sigma}E_{12}+n_{-\sigma}^2E_{11})\\
 \alpha_1 =&2 n_{-\sigma}\widetilde{n}E_{11}E_{12}- (\widetilde{n}+2n_{-\sigma})n_{-\sigma}E_{12}^2-n_{-\sigma}^3E_{11}^2\nonumber\\
& -E_{22}[E_{22} + 2n_{-\sigma}(n_{-\sigma}E_{11}-2E_{12})]\\
 \alpha_2 =&n_{-\sigma}^2(1-n_{-\sigma})^2E_{11}\\
%\mbox{and}\nonumber\\
 \alpha_3 =&n_{-\sigma}^2(1-n_{-\sigma})^2
\label{eqalfa}
\end{eqnarray} 
%\end{align}
%
and $\widetilde{n}=(1 + n_{-\sigma})$. The $E_{nm}$ are elements of the energy matrix (\ref{Enm}) and the quantity $P(\omega)$, is defined as:
\begin{eqnarray}
%\begin{align}
P(\omega)&=[(\omega-E_{11})(n_{-\sigma}\omega-E_{22})-(n_{-\sigma}\omega-E_{12})^2]\nonumber\\
         & \times[(\omega+E_{11})(n_{-\sigma}\omega+E_{22})-(n_{-\sigma}\omega+E_{12})^2]~~~\nonumber\\
         & +\Delta_{\vec{k}}^2(A_1-A_2\omega)
\label{eqPw}
\end{eqnarray} 
%\end{align}
%
with
\begin{equation}
A_1=a_0+a_1\frac{U}{\theta} + \left[a_2+\Delta_{\vec{k}}^2(1+n_{-\sigma}^2\frac{U}{\theta})^2\right]\left(\frac{U}{\theta}\right)^2
\end{equation}
and
\begin{equation}
A_2=(1+n_{-\sigma}^2\frac{U}{\theta})^2+n_{-\sigma}^2(1-n_{-\sigma})^2\left(\frac{U}{\theta}\right)^2.
\end{equation}

The quantities $a_0$, $a_1$, $a_2$ and $\theta$, are given by:
%
%\begin{align}
\begin{eqnarray}
a_0&= E_{11}^2\\
a_1&= 2E_{12}(2n_{-\sigma}E_{11}-E_{12}) \\
a_2&=E_{22}^2 - 4 n_{-\sigma}E_{12} E_{22}  + 2n_{-\sigma}^2(E_{12}^2 + E_{11} E_{22})
%\end{align}
\end{eqnarray}
and
\begin{equation}
\theta=tn_{01\sigma}-U(D_{01\sigma}+2\langle S_1^zS_0^z\rangle)
\label{eqtheta}
\end{equation} 
with the correlation functions $n_{01\sigma}$, $D_{01\sigma}$ and $\langle S_1^zS_0^z\rangle$ defined in the Appendix \ref{app:B}.

The main reason that we are adopting the d-wave symmetry is that we are following reference \cite{Edwards95} where it is claimed that for a large number of HTSC material d-wave gap symmetry is the most relevant.

Moreover, the d-wave symmetry follows also from the fact that in our Hamiltonian we consider an attractive delocalized interaction term. Actually the extended s-wave symmetry is more favoured for an attractive local interaction as discussed in \cite{bib1}.

For the particular case of pairing with $d$-wave symmetry, the gap function is
\begin{equation}
\Delta_{\vec{k}}=2\Delta[cos(k_x)-cos(k_y)],
\end{equation}
where $\Delta$ is the gap function amplitude.
Following the procedure described in references \cite{Edwards95,Eleonir2005}, the self-consistent gap function 
has been obtained from the Green's function: 
\begin{equation}
 G_{\sigma}^{dd^{\dagger}}(\vec{k},\omega)=-\frac{\Delta_{\vec{k}}(\beta_0 +\beta_1\omega^2)}{P(\omega)}
\label{eqG13}
\end{equation} 
in which,
\begin{equation}
\Delta=-2\theta\Delta\frac{1}{L}\sum_{\vec{q}}[cos(q_x)-cos(q_y)]^2F_{1\vec{q}\sigma}
\end{equation}
and
\begin{equation}
F_{1\vec{q}\sigma}=\frac{1}{2\pi i}\oint f(\omega)[\frac{\beta_0 +\beta_1\omega^2}{P(\omega)}]d\omega .
\end{equation}

The $\beta_0$ and $\beta_1$ are

\begin{eqnarray}
 \beta_0&=&n_{-\sigma}^2(1-n_{-\sigma})^2\frac{U}{\theta}\\
 \beta_1 &=&E_1^2-[E_2^2
-2E_1E_2+\Delta_{\vec{q}}^2(1+n_{-\sigma}^2\frac{U}{\theta})^2]\frac{U}{\theta}~~~~~~
\label{eqbeta}
\end{eqnarray} 
with $E_1=E_{12}-n_{-\sigma}E_{11}$ and $E_2=E_{22}-n_{-\sigma}E_{12}$.

\section{Self-consistent results and conclusions}
\label{sec:IV}

Firstly, we discuss the numerical results for the normal state ($T>T_c$).
\begin{figure}
\includegraphics[angle=-90,width=8cm]{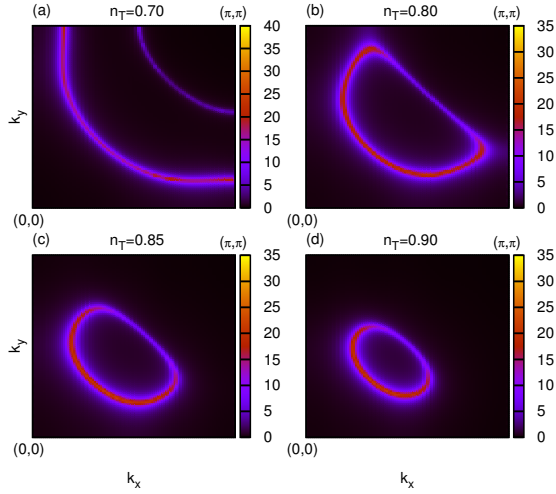}
\caption{(Color)The spectral function $A(\vec{k},\omega = 0)$ representing the Fermi surface for different dopings $\delta=1-n_T$. The model parameters considered here are $U=8t$, $t=-1.0$ eV, $t_{2}=0.3|t|$ and $k_BT=0.1|t|$ ($k_{B}$ is the Boltzmann constant).}
 \label{figFS}
\end{figure}
\begin{figure}
\includegraphics[angle=-90,width=8 cm]{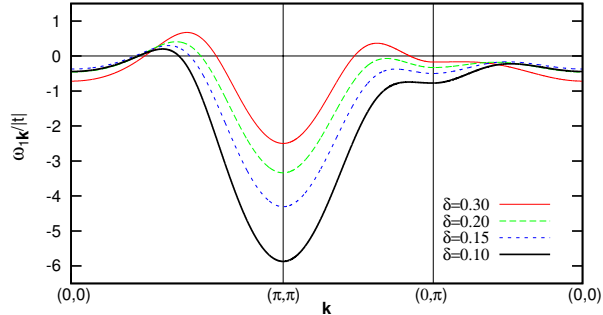}
\caption{(Color) The quasiparticle bands intercepted by the chemical potential ($\mu=0$). The model parameters and the temperature are the same as in figure \ref{figFS}.  }
\label{fig2}
\end{figure}
Figure \ref{figFS} shows the FS for four different doping values $\delta$. In (a), $\delta=0.30$, 
a well defined electron-like FS is found. 
However, in (b) when $\delta$ is decreased to
$\delta=0.20$, the topology of the FS changes, 
with the emergence of a hole-pocket enclosing the nodal point $(\frac{\pi}{2},\frac{\pi}{2})$. 
As a consequence, due to low spectral intensity, a pseudogap appears near the antinodal points 
$(\pi,0)$ and $(0,\pi)$. Further decrease in $\delta$ intensifies the presence of the pseudo-gap
as shown in (c)-(d).
This unusual behaviour is due to the strong antiferromagnetic correlations coming from the band shift $W_{\vec{k}\sigma }^d$ \cite{Edwards95,Eleonir2010} defined in equation (\ref{Apx32}).

The effects described above, are corroborated  by
the quasiparticle band calculation exhibited in figure \ref{fig2} which displays the quasiparticle band for distinct doping values $\delta$. 
In fact, while in the higher doping
regime the quasiparticle bands cross the Fermi level 
near $(\frac{\pi}{2},\frac{\pi}{2})$ and near the antinodal point $(0,\pi)$, in the lower doping
regime the quasiparticle band crosses the Fermi level only near the nodal point $(\frac{\pi}{2},\frac{\pi}{2})$. Such a behavior gives rise to a pocket around $(\frac{\pi}{2},\frac{\pi}{2})$. On the other hand, as the quasiparticle band does not touch the Fermi level near $(0,\pi)$, a pseudogap emerges 
in that region. 
As far as we know, the emergence of the pseudogap due to an attractive $U$ in a strong correlation regime, is for the first time presented here.
The kink observed near the $(\pi,\pi)$ point of the quasiparticle band is caused by
the strong antiferromagnetic correlations associated with $\langle \vec{S_j}\cdot\vec{S_i}\rangle$, 
which are maximum in {\bf Q}$=(\pi,\pi)$. The {\bf Q} is the antiferromagnetic wave-vector.

\begin{figure}[h]
\includegraphics[width=5.2cm,angle=-90]{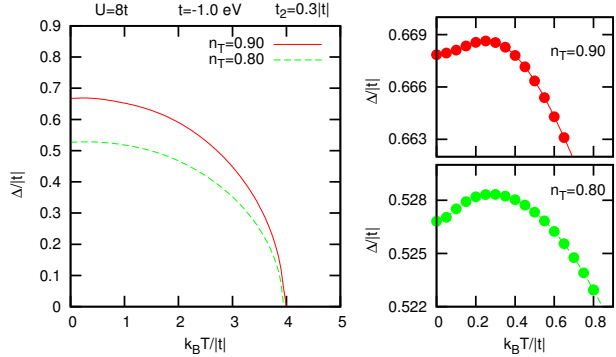}
 \caption{(Color)The main figure shows the gap function amplitude versus the temperature for $U=8t$, $t=-1.0$ eV, $t_{2}=0.3|t|$ and two different occupations $n_T$. The small figures show the regions of low temperatures where the gap presents an unusual behavior.   }
 \label{figgapant}
\end{figure}
Now we discuss some thermodynamical properties associated to the superconducting state. 
In figure \ref{figgapant} we describe the gap function amplitude $\Delta(T)$ for $U=8t$ and two different occupations $n_T$, in the lower doping 
regime. One sees that for a given $U$ (in a characteristic strong coupling regime $|\frac{U}{t}|>>1$), the zero temperature gap for $n_T=0.90$ is higher than the corresponding one for $n_T=0.80$.  Furthermore, the temperature where a non-superconducting phase arises is higher for $n_T=0.90$, i.e., $T_c(n_T=0.90)>T_c(n_T=0.80)$. 
It should be noted that in both cases for $n_T$, in the region of low $T$, there is a increase in the value of the gap amplitude as compared to the zero gap amplitude value. This unusual behavior is due to the effect of the strong correlations, since that in the BCS weak correlated regime, $\Delta(0)$ is always greater than $\Delta(T)$.

We stress that in our case this non-monotonic behavior at low temperatures is mainly due to correlation functions in the pole structure of the superconduction Green's function  (see equations (\ref{eqGS11}) and (\ref{eqG13})). To be more precise, we have found in our self-consistent calculation a complex interplay between the SC gap behaviour and the AF type short range correlations. 

\begin{figure}[t]
\includegraphics[width=5.2cm,angle=-90]{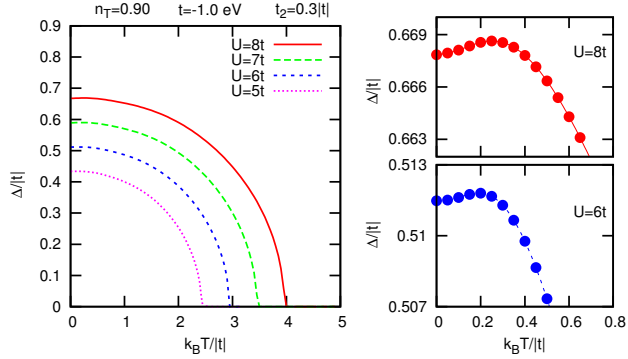}
\caption{(Color)The main figure shows the gap function amplitude versus the temperature for $n_T=0.90$ and different values of $U$. The remaining parameters are identical to figure \ref{figgapant}. The small figures show the regions of low temperatures for two values of $U$. For low temperatures, the gap presents an unusual behavior.}
 \label{figgapTant0.9}
\end{figure}

Figure \ref{figgapTant0.9} displays the value of the gap amplitude $\Delta(T)$ as a function of temperature for several values of $U$, in the strong correlation regime ($|\frac{U}{t}|>>1$). In all cases, for $n_T=0.90$, we note that when $U$ increases, $T_c$ increases also, and the same unusual behavior for $\Delta(T)$ appears for very low $T$. We have calculated the gap function amplitude for several values of $U$, for different $n_T$ ($n_T=0.80$ and 0.70), and the same behavior is observed.
\begin{figure}[h]
\includegraphics[width=6cm,angle=-90]{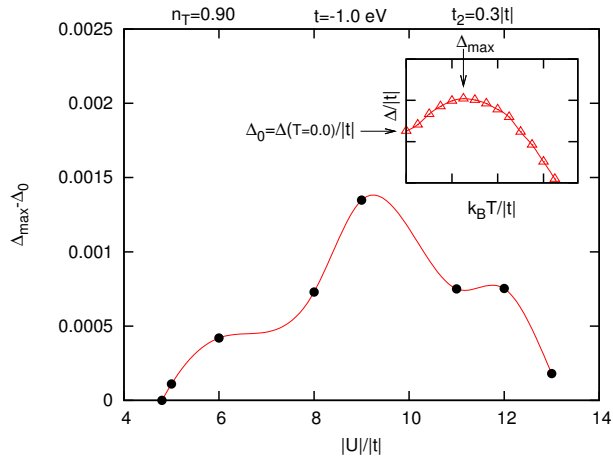}
\caption{(Color)The difference between $\Delta_{max}$ and $\Delta_{0}$ as a function of $|\frac{U}{t}|$. The inset shows as $\Delta_{max}$ and $\Delta_{0}$ have been obtained.}
 \label{figDeltagap}
\end{figure}

A very interesting result is shown in figure \ref{figDeltagap}. 
Here, we plot $\Delta_{max}$ - $\Delta_{0}$ for several values of $|\frac{U}{t}|$. One noted that when $U$ increases, $\Delta_{max}$ - $\Delta_{0}$ increases also, until a special value (in our calculation $|\frac{U}{t}|\simeq 8$) where $\Delta_{max}$ - $\Delta_{0}$ is a maximum and then, $\Delta_{max}$ - $\Delta_{0}$ tends toward zero in a very high value of $U$.
When such behavior appears, one has attained the $|U|\rightarrow \infty$ limit. Moreover, it should be noted that this unusual behavior occurs for a critical low value of $|U|$, (in our calculation $|\frac{U}{t}|\simeq 4$), which is a signature of the onset of a characteristic $|\frac{U}{t}|>> 1$, signaling the appearance of a strong correlation regime.

%Finally, 
Our results are in qualitatively agreement concerning the Fermi surface in underdoping region with other approaches using the t-J model \cite{tJ}. 
The reason of such agreement is that in both approaches the spin spin correlation functions renormalize the band structure giving rise to the appearance of hole-pockets.

Finally, in order to complement our present calculations, we need to discuss the higher doping
regime $n_T\lesssim 0.80$. Moreover, a detailed discussion of the effect of external pressure, which affects mainly the ratio $\frac{t_2}{t}$ \cite{angi}, is needed. These further calculations, are now in progress.

\appendix
\section{Two-pole approximation}
\label{app:A}

In the present two-pole approximation \cite{Roth69,Edwards95}, the Green's functions are defined as:
\begin{equation}
{\bf G}\left( \omega \right) ={\bf N}(\omega {\bf N}-{\bf E})^{-1}{\bf N}
\label{eq2.7}
\end{equation}
where, {\bf E} and {\bf N} are the energy and the normalization matrices given by
\begin{equation}
E_{nm}=\left\langle\left[ \left[ A_{n},H\right] _{-},A_{m}^{\dag}\right] _{\left( +\right) }\right\rangle
\label{Enm}
\end{equation}
and
\begin{equation}
N_{nm}=\langle [ A_{n},A_{m}^{\dag}]_{\left( +\right) }\rangle.
\label{Nnm}
\end{equation}

In equations (\ref{Enm})-(\ref{Nnm}), $[...,...]_{(+)-}$ denote the (anti)commutator, and $\langle ...\rangle$, the thermal average.  The set of operators $\{A_n\}$ must satisfy, within some approximation,
the relation  $\left[ A_{n},H\right] _{-}=\sum_{m}K_{nm}A_{m}$. 

For the normal state of the model (\ref{eqH}), the energy matrix
is given by, 
\begin{eqnarray}
{\bf E}=\left[
\begin{tabular}{cc}
$\varepsilon_{\vec{k}}-\mu +U_1 ~~~~~~ $     &
$\varepsilon_{\vec{k}}-\mu +U_2$\\
\\$\varepsilon_{\vec{k}}-\mu +U_2~~~~~~$     &
$ \varepsilon_{\vec{k}}n_{\sigma}^2-\mu+U_2+ \overline{n}W_{\vec{k}\sigma}$
\end{tabular}
\right]
%\nonumber \\
\label{eq2.10}
\end{eqnarray}
with
\begin{equation}
%\begin{align}
U_1=2U\sum_l\langle n_{l,-\sigma}\rangle,
\label{U_1}
%\end{align}
\end{equation}
\begin{equation}
%\begin{align}
U_2=2U\sum_l\langle n_{l,-\sigma}n_{i,-\sigma}\rangle
\label{U_2} 
%\end{align}
\end{equation}
and $\overline{n}=n_{-\sigma}(1-n_{-\sigma})$. The correlation function $D_{il-\sigma}=\langle n_{i-\sigma}n_{l-\sigma}\rangle$ is defined in equation (\ref{eqD01}).
The band shift $W_{\vec{k}\sigma}$, can be written as:
\begin{equation}
\overline{n}W_{\vec{k}\sigma }^d=-\sum_{\langle\langle j\neq 0\rangle\rangle}t_{0j}(n_{0j\sigma } - 2m_{0j\sigma })
+\sum_{\langle\langle j\neq 0\rangle\rangle}t_{0j}e^{i\vec{k}\cdot\vec{R}_j}h_{j\sigma}
\label{Apx32}
\end{equation}
with $n_{ij\sigma}$, $m_{ij\sigma }$ and $h_{j\sigma}$ defined below. 

\section{Correlation functions}
\label{app:B}

The correlation function $n_{ij\sigma}=\langle d_{i\sigma}^{\dagger}d_{j\sigma}\rangle$, is given by:
\begin{equation}
n_{ij\sigma}=\frac{1}{2\pi i L}\sum_{\vec{k}}\oint e^{i\vec{k}\cdot(\vec{R}_j-\vec{R}_i)} f(\omega)G_{S\sigma}^{dd}(\vec{k},\omega)d\omega
\label{eqnij}
\end{equation}
with $G_{S\sigma}^{dd}$ defined in equation (\ref{eqGS11}).
Assuming $i=0$ and $t_{0j}=t$ for the $z$ nearest-neighbors, only one value of $n_{0j\sigma}$, namely $n_{01\sigma}$, is necessary. Considering the same for the second-nearest-neighbors $t_2$, we have:
\begin{equation}
n_{01\sigma}=\frac{1}{2\pi i L}\sum_{\vec{k}}\oint \frac{\epsilon_{\vec{k}}}{(t+t_2)z}f(\omega)d\omega .
\label{eqn1}
\end{equation}

By using the original Roth's scheme \cite{Roth69}, the correlation function $D_{ij\sigma}=\langle n_{i\sigma}n_{j\sigma}\rangle$, is calculated and written as:
\begin{equation}
D_{ij\sigma}=n_{\sigma}^2-\frac{\alpha_{ij\sigma}n_{ij\sigma}+\beta_{ij\sigma}m_{ij\sigma}}{1-\beta_{ii,\sigma}\beta_{ii,-\sigma}}
\label{eqD01}
\end{equation}
with $m_{ij\sigma}=\langle d_{i\sigma}^{\dagger}n_{j-\sigma}d_{j\sigma}\rangle$ given by
\begin{equation}
m_{ij\sigma}=\frac{1}{2\pi i L}\sum_{\vec{k}}\oint e^{i\vec{k}\cdot(\vec{R}_j-\vec{R}_i)} f(\omega)G_{S\sigma}^{n_2d}(\vec{k},\omega)d\omega 
\label{eqmij}
\end{equation}
and 
\begin{equation}
%\begin{align}
\alpha_{ij\sigma}=\frac{n_{ij\sigma}-m_{ij\sigma}}{1-n_{-\sigma}}
\label{eqalfa2}
%\end{align}
\end{equation}
%
%
%\begin{align}
\begin{equation}
\beta_{ij\sigma}= \frac{m_{ij\sigma}/n_{-\sigma}-n_{ij\sigma}}{1-n_{-\sigma}} .
\label{beta2}
\end{equation}
%\end{align}
The $D_{01\sigma}$ can be obtained assuming again $i=0$ and $t_{0j}=t$ for the $z$ nearest-neighbors, as it has been done in $n_{01\sigma}$.

The Green's function $G_{S\sigma}^{n_2d}$ in \ref{eqmij}, is:
\begin{equation}
G_{S\sigma}^{n_2d}(\vec{k},\omega)=\frac{n_{-\sigma}\left[A''(\omega)-A'''(\omega)(1+n_{-\sigma}^2\frac{U}{\theta})\Delta_{\vec{k}}^2\right]}{P(\omega)}
\label{eqGnd}
\end{equation} 
where
\begin{equation}
A''(\omega)=\gamma_0+ \gamma_1\omega+ \gamma_2\omega^2+ \gamma_3\omega^3
\label{eqAl}
\end{equation} 
with
\begin{eqnarray}
%\begin{align}
 \gamma_0=&(E_{12}^2-E_{11}E_{22})[E_{3}+n_{-\sigma}(E_{12}-E_{11})]\\
 \gamma_1 =&n_{-\sigma}E_{11}[E_{12}(1+3n_{-\sigma})-n_{-\sigma}(E_{11}+\widetilde{n}E_{22})]\nonumber \\
&+E_{22}E_{3}+n_{-\sigma}E_{12}(3E_{3}-n_{-\sigma}E_{12})\\
\gamma_2 =&n_{-\sigma}(1-n_{-\sigma})^2E_{12}\\
 \gamma_3 =&n_{-\sigma}^2(1-n_{-\sigma})^2
\label{eqgamma}.
\end{eqnarray}
%\end{align}
%
The quantities $E_3$ and $\widetilde{n}$ are:
\begin{equation}
%\begin{align}
 E_3=E_{22}-E_{12}~~~~\mbox{and}~~~~\widetilde{n}=1+n_{-\sigma}.
%\end{align}
\end{equation}

The term $A'''$ introduced in equation (\ref{eqGnd}) is defined as: 
\begin{equation}
A'''(\omega)=\omega(1+n_{-\sigma}^2\frac{U}{\theta})+ E_{11}+\frac{U}{\theta}[n_{-\sigma}(E_{12}+E_{11})-E_{12}].
\label{eqAlll}
\end{equation} 
The denominator of the Green's function $G^{n_2d}$, is given in equation (\ref{eqPw}).

The term $h_{j\sigma}$ presented in the band shift (\ref{Apx32}), is given by:
\begin{equation}
h_{j\sigma}= B_{j\sigma} + \langle \vec{S_j}\cdot\vec{S_0}\rangle 
\label{h2}
\end{equation}
with
%
%\begin{align}
\begin{eqnarray}
B_{j\sigma}&=&-\langle S_j^zS_0^z \rangle-\frac{\alpha_{j\sigma}n_{0j\sigma}^d + \beta_{j\sigma}m_{j\sigma}}{1-\beta_{\sigma}\beta_{-\sigma}} \nonumber\\ 
& &-\frac{\alpha_{j\sigma}n_{0j-\sigma}^d +\beta_{j\sigma}(n_{0j-\sigma}^d
-m_{j-\sigma} ) }{1-\beta_{\sigma}}
\label{Ajs}
%\end{align}
\end{eqnarray}
and
\begin{equation}
\langle \vec{S_j}\cdot\vec{S_0}\rangle= \frac{1}{2}\left(\langle S_j^+S_0^-\rangle 
+\langle S_j^-S_0^+\rangle\right) +\langle S_j^zS_0^z \rangle 
\label{SjSi}.
\end{equation}
Particularly, in the paramagnetic state, $\langle S_j^+S_0^-\rangle=\langle S_j^-S_0^+\rangle$,
then, $\langle \vec{S_j}\cdot\vec{S_0}\rangle$ can be written as:
\begin{equation}
\langle \vec{S_j}\cdot\vec{S_0}\rangle= \langle S_j^+S_0^-\rangle +\langle S_j^zS_0^z \rangle 
\label{SjSi2}
\end{equation}
where,
\begin{equation}
\langle S_j^+S_0^- \rangle =\langle d_{j\sigma}^{\dagger}d_{j-\sigma}d_{0-\sigma}^{\dagger}d_{0\sigma} \rangle=-\frac{\alpha_{j\sigma}n_{0j,-\sigma}^d+\beta_{j,\sigma}m_{j,-\sigma} }{1+\beta_{\sigma}}
\label{S+S-}
\end{equation}
and
\begin{eqnarray}
%\begin{align}
\langle S_j^zS_0^z \rangle &=& \frac{(1-\beta_{-\sigma})}{2}\left[(n_{\sigma}^d)^2-\frac{\alpha_{j\sigma}
n_{0j\sigma}+\beta_{j\sigma}m_{j\sigma} ) }{1-\beta_{\sigma}\beta_{-\sigma}}\right] 
%\nonumber\\ 
-\frac{\alpha_{-\sigma}n_{\sigma}^d}{2} 
\label{SzSz}.
%\end{align}
\end{eqnarray}
The correlation functions $n$, $m$, $\alpha$ and $\beta$, are defined in equations (\ref{eqnij}), (\ref{eqmij}), (\ref{eqalfa2}) and (\ref{beta2}), respectively.

In order to calculate $n_{ij\sigma}$ and $m_{ij\sigma}$ in the normal state, it is necessary to consider $\Delta =0$ in the Green's functions $G_{S\sigma}^{dd}$ and $G_{S\sigma}^{n_2d}$ defined in equations (\ref{eqGS11}) and (\ref{eqGnd}). In this case, $G_{S\sigma}^{dd}\rightarrow G_{N\sigma}^{dd}$ and $G_{S\sigma}^{n_2d}\rightarrow G_{N\sigma}^{n_2d}$.

\end{document}